\documentclass[useAMS,usenatbib]{mn2e}

\usepackage{epsfig}
\usepackage{amsmath}
\usepackage{amssymb}
\usepackage{natbib}
\usepackage{threeparttable} 
\usepackage{booktabs}

\usepackage{epstopdf}
\usepackage{times}

\newcommand{\swift}{\textit{Swift}}

\newcommand{\maxi}{\textit{MAXI}}

\newcommand{\nustar}{\textit{NuSTAR}}

\newcommand{\Msun}{\mathrm{M}_{\odot}}
\newcommand{\lum}{\mathrm{erg~s}^{-1}}
\newcommand{\flux}{\mathrm{erg~cm}^{-2}~\mathrm{s}^{-1}}

\newcommand{\cnts}{\mathrm{counts~s}^{-1}}

\newcommand{\nh}{\mathrm{cm}^{-2}}

\newcommand{\gmc}{GM/c^2}
\newcommand{\risco}{R_{\mathrm{ISCO}}}
\newcommand{\relx}{\textsc{relxilllp}}

\newcommand{\hete}{HETE J1900.1--2455}

\newcommand{\nsgro}{GRO~J1744-28}

\newcommand{\source}{4U 1608--52}

\def \mnras {MNRAS}
\def \apj {ApJ}
\def \apjs {ApJS}
\def \apss {Ap\&SS}
\def \apjl {ApJL}
\def \aap {A\&A}

\def \atel {The Astronomer's Telegram}

\title[\nustar\ observation of 4U 1608--52]{A \nustar\ observation of disc reflection from close to the neutron star in 4U 1608--52}
\author[N. Degenaar et al.]
{N. Degenaar$^1$\thanks{e-mail: degenaar@ast.cam.ac.uk}, J.M. Miller$^{2}$, D. Chakrabarty$^{3}$, F.A. Harrison$^{4}$, E.~Kara$^1$ and A.C.~Fabian$^1$ \\ 
$^1$Institute of Astronomy, University of Cambridge, Madingley Road, Cambridge CB3 OHA, UK\\
$^2$Department of Astronomy, University of Michigan, 1085 South University Avenue, Ann Arbor, MI  48109, USA\\
$^3$Massachusetts Institute of Technology (MIT), Kavli Institute for Astrophysics and Space Research, Cambridge, MA 02139, USA\\
$^4$Cahill Center for Astronomy and Astrophysics, California Institute of Technology, Pasadena, CA, 91125 USA 
}

\begin{document}

\date{Accepted 2015 May 8. Received 2015 May 8; in original form 2015 April 2}

\pagerange{\pageref{firstpage}--\pageref{lastpage}} \pubyear{0000}

\maketitle

\label{firstpage}

\begin{abstract}
Studying the reflection of X-rays off the inner edge of the accretion disc in a neutron star low-mass X-ray binary, allows us to investigate the accretion geometry and to constrain the radius of the neutron star. We report on a \nustar\ observation of \source\ obtained during a faint outburst in 2014 when the neutron star, which has a known spin frequency of $\nu=620$~Hz, was accreting at $\simeq$1--2 per cent of the Eddington limit. The 3--79 keV continuum emission was dominated by a $\Gamma \simeq 2$ power law, with an $\simeq$1--2 per cent contribution from a $kT_{\mathrm{bb}}$$\simeq$0.3--0.6~keV blackbody component. The high-quality \nustar\ spectrum reveals the hallmarks of disc reflection; a broad iron line peaking near 7~keV and a Compton back-scattering hump around $\simeq$20--30 keV. Modelling the disc reflection spectrum points to a binary inclination of $i\simeq$30$^{\circ}$--40$^{\circ}$ and a small `coronal' height of $h\lesssim$8.5$~\gmc$. Furthermore, our spectral analysis suggests that the inner disc radius extended to $R_{\mathrm{in}}$$\simeq$7--10$~\gmc$, close to the innermost stable circular orbit. This constrains the neutron star radius to $R\lesssim$21~km and the redshift from the stellar surface to $z \gtrsim$0.12, for a mass of $M = 1.5~\Msun$ and a spin parameter of $a=0.29$.
\end{abstract}



\section{Introduction}
Measuring the mass and radius of neutron stars would constrain their equation of state and thereby give insight into the behaviour of matter at supranuclear densities, a challenge that cannot be achieved in terrestrial laboratories. Since many proposed equations of state predict a range of masses for a given radius \citep[see e.g.,][for a recent review]{lattimer2011}, it is of particular interest to measure neutron star radii. Low-mass X-ray binaries (LMXBs) are promising targets to obtain such information, as the hot surface of the neutron star may be directly visible in the X-ray band. 

When residing in an LMXB, a neutron star accretes matter from an accretion disc that is fed by a low-mass ($\lesssim1~\Msun$) companion star. Often LMXBs are transient, exhibiting orders of magnitude variation in their X-ray luminosity driven by similarly large variations in the mass-accretion rate. During quiescent episodes the X-ray luminosity is low ($L_{\mathrm{X}}\simeq10^{32-34}~\lum$) and little matter is thought to be accreted on to the neutron star. Thermal emission from the stellar surface may then be detected with sensitive X-ray instruments and used to measure neutron star radii \citep[e.g.,][]{rutledge1999,webb2007,guillot2011,guillot2013,servillat2012,heinke2006,heinke2014}. 

During outburst phases, on the other hand, matter is rapidly accreted on to the neutron star, typically generating an X-ray luminosity of $\simeq$1--100 per cent of the Eddington limit \citep[$L_{\mathrm{Edd}}\simeq3.8\times10^{38}~\lum$;][]{kuulkers2003}. Although the overall X-ray emission is then dominated by that of the accretion flow, the neutron star becomes visible during type-I X-ray bursts; bright flashes of X-ray emission resulting from unstable thermonuclear burning of accreted H/He on the surface of the neutron star. Modelling the black-body spectra of these events also facilitates radius measurements \citep[e.g.,][]{vanparadijs1979,fujimoto1989_1608,guver2010,suleimanov2011_eos,poutanen2014_4u1608}.

Both these methods rely on using appropriate neutron star atmosphere models, which has spurred intense discussion \citep[e.g.,][]{suleimanov2011_eos,heinke2014}. Alternative means of constraining neutron star radii are offered by studying X-rays reflected off the inner edge of the accretion disc, which may extend very close to the stellar surface \citep[e.g.,][]{cackett2010_iron,miller2013_serx1}. The observable effects of disc reflection are a broad emission line in the Fe-K band (6.4--6.97 keV) and a Compton back-scattering hump peaking at $\simeq$20--40~keV \citep[][]{fabian1989}.

Studying reflection spectra also offers valuable insight into the accretion geometry, such as the inner radius and inclination of the accretion disc as well as the height of the illuminating X-ray source, and how this is affected by the accretion rate or the magnetic field of the neutron star. Analysis of broad Fe lines in several neutron star LMXBs has revealed inner disc radii of $R_{\mathrm{in}}$$\simeq$5--20$~\gmc$ \citep[e.g.,][for recent studies]{cackett2010_iron,egron2011,sanna2014,disalvo2015}, i.e., close to the innermost stable orbit (ISCO) in the Schwarzschild metric ($\risco$$\simeq$$6~\gmc$). However, in some LMXBs the disc appears to be truncated at larger radii due to the pressure exerted by the magnetic field of the neutron star \citep[e.g., \nsgro\ with $R_{\mathrm{in}}$$\simeq$85$~\gmc$;][]{degenaar2014_groj1744}, or due to evaporation of the inner disc at low accretion rates \citep[e.g., \hete\ with $R_{\mathrm{in}}$$\simeq$25$~\gmc$;][]{papitto2013_hete}. 

Detecting a Compton hump in addition to a broad iron line requires high sensitivity at energies $>$10~keV, such as provided by the recently launched \nustar\ satellite \citep[][]{harrison2013_nustar}. High quality \nustar\ spectra can allow for new views of the accretion geometry in LMXBs \citep[e.g.,][]{miller2015}, and constraints of neutron star radii \citep[e.g., Ser X-1;][]{miller2013_serx1}.

\subsection{\source}
In this work we report on the disc reflection spectrum measured by \nustar\ for the neutron star \source\ \citep[][]{grindlay1976,tananbaum1976}. This transient LMXB is frequently active, with accretion outbursts typically recurring once every $\simeq$1--2~yr \citep[e.g.,][]{lochner1994,chen97,simon2004,galloway06}. Type-I X-ray bursts are regularly observed and have allowed for a distance estimate of $D\simeq2.9-4.5$~kpc \citep[e.g.,][]{galloway06,poutanen2014_4u1608}. Moreover, rapid oscillations detected during type-I X-ray bursts revealed that the neutron star spins at $\nu=620$~Hz \citep[][]{muno2001,galloway06}. Renewed activity was detected from the source on 2014 October 5 with \maxi\ \citep[][]{negoro2014_4u1608}.

\begin{figure}
 \begin{center}
\includegraphics[width=7.9cm]{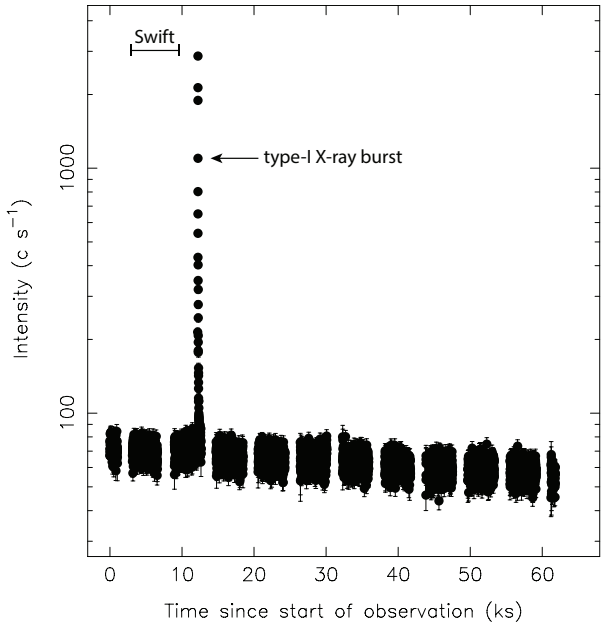}
    \end{center}
\caption[]{{\nustar\ FMPA/FMPB summed, background-corrected light curve at 1-s resolution (3--79 keV). The time of the \swift/XRT observation is indicated by the horizontal bar.}}
 \label{fig:lc}
\end{figure}


\section{Observations and analysis}
We observed \source\ simultaneously with \nustar\ and \swift\ \citep[][]{gehrels2004} on 2014 October 16--17. Data reduction and analysis was carried out using tools incorporated in \textsc{heasoft} ver. 16.6. Throughout this work we assume a distance of $D$$=$3.6~kpc and report errors as 90 per cent confidence levels.

\subsection{\nustar}
\nustar\ observed \source\ between 23:00 \textsc{ut} on 2014 October 16 and 17:10 \textsc{ut} on October 17 (ID 90002002002). Standard screening and processing with \textsc{nustardas} (ver. 1.4.1) resulted in $\simeq$32~ks on-target exposure time. We created light curves and spectra for the FPMA and FPMB employing the \textsc{nuproducts} tool. To obtain source events we used a circular extraction region with a radius of 120 arcsec, whereas a region of the same dimensions placed away from the source was used for the background.

Light curves obtained for the two modules were first background subtracted and then summed using \textsc{lcmath}. Initial inspection of the spectra revealed that the source was detected significantly above the background in the entire \nustar\ band (3--79 keV), and that the separate FPMA/FPMB data showed excellent agreement. We therefore combined these using \textsc{addascaspec}, which also generates the combined background spectrum and ancillary response file (arf). A combined redistribution matrix file (rmf) was created via \textsc{addrmf}, weighting the responses of the two modules by their exposure times. Using \textsc{grppha} we grouped the data to a minimum of 20 photons per spectral bin. 

\subsection{\swift}
Simultaneous \swift/XRT data were obtained to provide energy coverage down to $\simeq$0.5~keV. \source\ was observed for $\simeq$1.7~ks between 23:56 \textsc{ut} 2014 October 16 and 01:40 \textsc{ut} October 17 (ID 32322019), with the XRT operated in windowed timing (WT) mode. Using \textsc{xselect} we extracted source events from a box of 120 arcsec long and 40 arcsec wide. A box of the same dimensions placed away from the source was used to extract a background spectrum. 

The WT light curve showed a stable intensity along the observation ($\simeq$$5~\cnts$), and we therefore extracted a single average spectrum. An arf was created using \textsc{xrtmkarf} and the latest rmf (ver. 15) was sourced from the calibration data base. The spectral data were grouped to a minimum of 20 photons per bin.

\begin{figure}
 \begin{center}
\includegraphics[width=8.5cm]{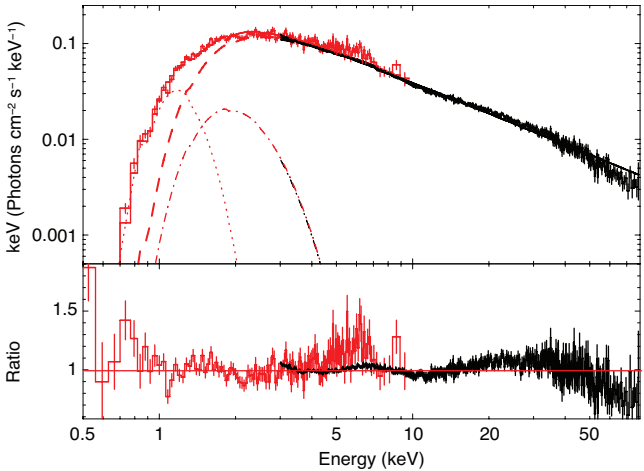} 
    \end{center}
\caption[]{{
Unfolded \nustar\ (black) and \swift\ (red) spectra. The solid lines indicate fits to an absorbed, phenomenological continuum consisting of a $\Gamma\simeq2.0$ power law (dashed lines), a cool $kT_{\mathrm{bb}}\simeq0.1$~keV blackbody (dotted curve), and a hotter $kT_{\mathrm{bb}}\simeq0.4$~keV blackbody (dash-dotted curves). The bottom panel shows the data-to-model ratio. 
}}
 \label{fig:contspec}
\end{figure}


\section{Results}\label{sec:results}

\subsection{\nustar\ light curve}\label{subsec:lcurve}
Fig.~\ref{fig:lc} shows the light curve of \source\ obtained with \nustar. The source was detected at an average intensity of $\simeq$$70~\cnts$ (3--79 keV, two modules combined). A strong $\simeq$200-s long increase in intensity was registered $\simeq$12~ks into the observation, caused by the occurrence of a type-I X-ray burst from the source. Analysis of this event will be presented in a separate work.

\subsection{Spectral continuum}\label{subsec:cont}
We examined the spectral continuum by fitting the \nustar\ data (excluding the X-ray burst) together with the \swift/XRT data in \textsc{xspec} \citep[ver. 12.8;][]{xspec}. A constant multiplication factor was included to account for calibration differences. To model the interstellar absorption we used \textsc{tbabs} with \textsc{vern} cross-sections \citep[][]{verner1996} and \textsc{wilm} abundances \citep[][]{wilms2000}. 

\begin{figure}
 \begin{center}
\includegraphics[width=8.3cm]{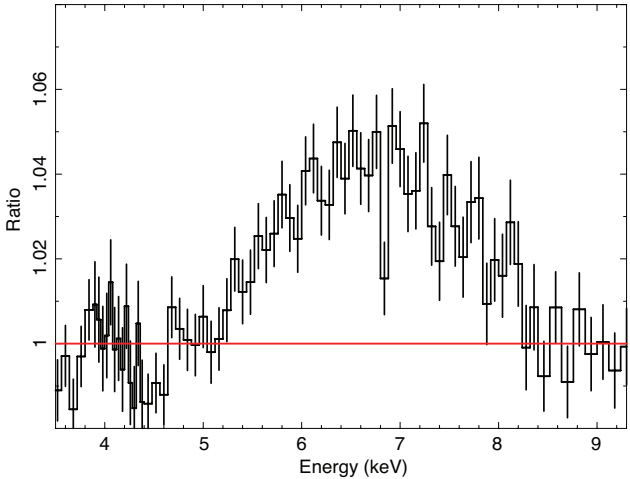} 
    \end{center}
\caption[]{{
\nustar\ data-to-model ratio in the Fe-K line region for a spectral continuum without disc reflection (rebinned for visual clarity). }}
 \label{fig:feline}
\end{figure}

Any realistic combination of continuum models left large positive residuals around $\simeq$5--8 and 20--30~keV. This is illustrated by Fig.~\ref{fig:contspec}, where we show a fit to a continuum consisting of a power law and two blackbody components. The prominent residuals can be interpreted as a broad Fe-K emission line, shown in more detail in Fig.~\ref{fig:feline}, and the corresponding Compton hump. We therefore proceeded by modelling our data with physical reflection models.

\subsection{Reflection spectrum}\label{subsec:reflection}
As can be seen in Fig.~\ref{fig:contspec}, the continuum emission at energies of $\gtrsim$2~keV is strongly dominated by a hard spectral component. We therefore chose the \relx\ model, which calculates disc reflection features due to an illuminating power-law source \citep[][]{garcia2014}. This new model features higher spectral resolution and updated atomic data with regard to other reflection models. Moreover, it combines the relativistic convolution kernel \textsc{relconv\_lp} \citep[][]{dauser2010} with the reflection grid \textsc{xillver} \citep[][]{garcia2013}, in which the reflection spectrum is calculated for each emission angle rather than averaged. Both the illuminating power law and the reflected emission are fitted self-consistently. 

In \relx\ the hard X-ray source, usually referred to as the corona, is assumed to be a point source located at a height $h$ above the accretion disc plane along the spin axis of the compact object. Although this `lamppost geometry' is an oversimplification (albeit compact, the corona is likely not a point source), it provides an adequate description for several LMXBs \citep[e.g.,][]{miller2013_BHreflection,miller2015}. The emissivity profile is calculated based on the height $h$ rather than assuming a (broken) power-law profile. 

The fit parameters of the \relx\ model are the dimensionless spin $a$, the binary inclination $i$, the inner and outer disc radii $R_{\mathrm{in}}$ and $R_{\mathrm{out}}$ (expressed in terms of $\risco$ for the given $a$), the ionization parameter $\log \xi$, the iron abundance $A_{\mathrm{Fe}}$ (with respect to Solar), the reflection fraction $R_{\mathrm{refl}}$ (ratio of the reflected to primary emission between 20 and 40 keV), the normalization $N_{\mathrm{refl}}$, and the  index $\Gamma$ and high-energy cutoff $E_{\mathrm{cut}}$ of the power law. Since \source\ has a known spin frequency, the dimensionless spin parameter can be calculated. For neutron stars this can be approximated as $a \simeq 0.47/P$[ms] \citep[][]{braje2000}. Given that $\nu = 620$~Hz \citep[][]{muno2001,galloway06}, we set $a=0.29$ in all our fits. Moreover, we fixed $R_{\mathrm{out}}=400~\risco$ since the emissivity profile drops off steeply with increasing radius, so that reflection fits are not sensitive to the outer disc radius.

Fitting the spectral data required one or two soft emission components in addition to the reflection model, for which we used the model \textsc{bbodyrad}. The thermal emission may originate from the neutron star surface, the accretion disc, or a boundary layer between the two. We found that the relatively high extinction towards \source\ and the limited number of counts in the \swift\ data caused considerable uncertainty in the spectral shape at $\lesssim$3~keV. To investigate to which extent this influenced the (reflection) spectrum at higher energies, we explored a number of different fits. These are summarized in Table~\ref{tab:spec} and discussed in more detail below.  

We initially fitted the combined \nustar\ and \swift\ data using the full 0.5--79 keV energy range, which required two soft components (model 1 in Table~\ref{tab:spec}). However, the obtained normalization of the $\simeq$0.1~keV blackbody would imply an unphysically large emission radius of $\gtrsim$$500~\gmc$. Possibly, the requirement for this soft component is due to a calibration uncertainty that may arise for highly absorbed sources observed in WT mode and causes a bump at energies of $\simeq$0.5--1 keV.\footnote{http://www.swift.ac.uk/analysis/xrt/digest$\_$cal.php} To mitigate these possible effects we also performed fits in the 1--79 keV range, which required only one blackbody component (model 2 in Table~\ref{tab:spec}). 

\begin{figure}
 \begin{center}
\includegraphics[width=8.5cm]{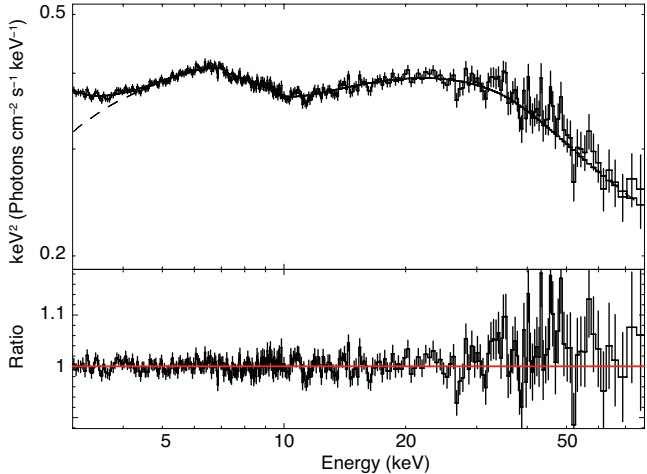}
    \end{center}
\caption[]{{Unfolded \nustar\ spectrum and model fit (number 4 in Table~\ref{tab:spec}; solid line) consisting of an illuminating power law and resulting disc reflection  (\relx; dashed line) plus a blackbody (\textsc{bbodyrad}; contributing only at $\lesssim$5~keV). The bottom panel shows the data-to-model ratio.}}
 \label{fig:reflspec}
\end{figure}

We also explored fits using only the \nustar\ spectral data. When leaving the hydrogen column density free (model 3 in Table~\ref{tab:spec}) the obtained value is notably larger than for fits 1--2, and also larger than the values of $N_{\mathrm{H}}\simeq (0.5-2)\times10^{22}~\nh$ reported previously \citep[e.g.,][]{penninx1989,keek2008_1608,guver2010}. We therefore also fitted the \nustar\ spectrum with $N_{\mathrm{H}} = 2\times10^{22}~\nh$ fixed (model 4 in Table~\ref{tab:spec}), shown in Fig.~\ref{fig:reflspec}. 

Importantly, we find that despite the considerably different parameter values that we obtain for the absorption and thermal emission component(s), the results obtained for the hard X-ray continuum and the reflection spectrum are robust (Table~\ref{tab:spec}). Our  fits point to a moderately low inclination of $i$$\simeq$30--40$^{\circ}$, consistent with observations of the optical counterpart (see Section~\ref{subsec:incl}). The obtained inner disc radius lies near the ISCO, $R_{\mathrm{in}}$$\simeq$1.3--2.0~$\risco$, and the coronal height is small, $h\lesssim$8.5~$\gmc$. We find a reflection fraction of $R_{\mathrm{refl}}$$\simeq$1 and a high ionization parameter of $\log \xi\simeq3.9$, as may be expected given the breadth of the Fe line (Fig.~\ref{fig:feline}). 

We note that our spectral fits favour a high value for the power-law cutoff; $E_{\mathrm{cut}}\gtrsim300$~keV. Although this is not unprecedented for neutron stars, it is well outside the \nustar\ bandpass. This could perhaps indicate that the spectral shape of the illuminating X-ray source (slightly) deviates from the cutoff power law that is incorporated in the \relx\ model.  

Our different fits yield an unabsorbed flux in the 3--79 keV band of $F_{3-79}\simeq$$2.0\times10^{-9}~\flux$, of which $\simeq$1--2 per cent is due to thermal emission. The uncertainty in the low-energy spectrum causes some spread in the inferred 0.5--79 keV fluxes; we obtain $F_{0.5-79}\simeq$$(3.0-5.2) \times 10^{-9}~\flux$ for fits 2--4, and $\simeq$$1.1 \times 10^{-8}~\flux$ for fit 1. The latter may be  overestimated due to a possible calibration issue in the WT data. We therefore estimate that \source\ accreted at $\simeq$1--2 per cent of the Eddington limit during our observations.

\begin{table*}
\caption{ Results from modelling the spectral data with disc reflection.}
\begin{threeparttable}
\begin{tabular*}{1.01\textwidth}{@{\extracolsep{\fill}}lcccc}
\hline
Model number & 1 & 2 & 3 & 4 \\
Fitted data & \swift+\nustar\ 0.5--79 keV & \swift+\nustar\ 1--79 keV  & \nustar\ 3--79 keV & \nustar\ 3--79 keV \\
\textsc{constant*tbabs*} & (\textsc{bbodyrad+bbodyrad+relxilllp}) & (\textsc{bbodyrad+relxilllp}) & (\textsc{bbodyrad+relxilllp}) & (\textsc{bbodyrad+relxilllp}) \\
\hline
$C$ \dotfill & $1.04 \pm 0.02$ & $1.02 \pm 0.01$ & -- & --  \\
$N_{\mathrm{H}}$ ($\times10^{22}~\nh$) \dotfill & $2.6 \pm 0.2$ & $1.7 \pm 0.1$ & $3.9 \pm 1.1$ & 2.0 fix \\
$kT_{\mathrm{bb1}}$ (keV)  \dotfill & $0.46 \pm 0.02$ & $0.56 \pm 0.03$  & $0.43 \pm 0.04$  & $0.31 \pm 0.03$ \\
$N_{\mathrm{bb1}}$ (km/10~kpc)$^2$  \dotfill & $6.0^{+2.5}_{-2.2} \times 10^2$ & $1.4^{+0.5}_{-0.4} \times10^2$ & $3.7^{+3.7}_{-1.5} \times 10^3$ & $1.1^{+2.2}_{-0.7} \times 10^4$\\
$kT_{\mathrm{bb2}}$ (keV)  \dotfill & $0.13 \pm 0.01$ & --  & -- & -- \\
$N_{\mathrm{bb2}}$ (km/10~kpc)$^2$  \dotfill & $9.9^{+1.2}_{-5.6} \times 10^6$ & -- & -- & -- \\
$h$ ($\gmc$)  \dotfill & $1.9^{+6.6}_{-1.9}$ & $1.9^{+3.3}_{-1.9}$ & $0.8^{+5.5}_{-0.8}$ & $1.9^{+6.6}_{-1.9}$ \\
$i$ ($^{\circ}$)  \dotfill & $36.1 \pm 6.9$ & $31.1 \pm 6.5$ & $41.1 \pm 6.1$ & $38.8 \pm 6.0$ \\
$R_{\mathrm{in}}$ ($\times \risco$)  \dotfill & $2.0^{+1.3}_{-0.6}$ & $1.7^{+0.6}_{-0.7}$ & $1.3^{+0.8}_{-1.3}$ & $1.9^{+1.9}_{-0.6}$  \\
$\Gamma$  \dotfill & $1.96 \pm 0.02$ & $1.97 \pm 0.01$ & $1.97 \pm 0.01$ & $1.95\pm0.01$ \\
$\log \xi$   \dotfill & $4.0 \pm 0.2$ & $4.1 \pm 0.1$ & $3.9 \pm 0.2$ & $3.9\pm0.1$ \\
$A_{\mathrm{Fe}}$ ($\times$~Solar)  \dotfill & $2.2 \pm 0.5$ & $2.5 \pm 0.7$ & $1.7 \pm 0.7$ & $1.8\pm0.4$ \\
$E_{\mathrm{cut}}$ (keV)  \dotfill & $439^{+101}_{-77}$ & $554^{+100}_{-150}$ & $513^{+254}_{-197}$ & $416^{+110}_{-58}$ \\
$R_{\mathrm{refl}}$  \dotfill & $1.3 \pm 0.7$  & $1.5 \pm 0.5$  & $1.0 \pm 0.4$  & $1.0 \pm 0.3$ \\
$N_{\mathrm{refl}}$   \dotfill & $0.15 \pm 0.03$  & $0.14 \pm 0.03$  & $0.18 \pm 0.03$  & $0.18\pm0.03$ \\
$\chi_{\nu}^2$ (dof)  \dotfill & 1.18 (1847) & 1.22 (1811) & 1.06 (1280) & 1.07 (1281) \\
\hline
\end{tabular*}
\label{tab:spec}
\begin{tablenotes}
\item[] Notes. -- The constant $C$ was fixed to 1 for the \nustar\ data and left free for the \swift\ spectrum. The outer disc radius of the \relx\ spectral component was always fixed at $R_{\mathrm{out}}=400~\risco$ and the spin parameter was set to $a=0.29$ to correspond to the known spin frequency of $\nu=620$~Hz. The reflection fraction $R_{\mathrm{refl}}$ is calculated between 20 and 40~keV. Quoted errors reflect 90 per cent confidence levels. 
\end{tablenotes}
\end{threeparttable}
\end{table*}

\section{Discussion}\label{sec:discussion}
The frequently active transient neutron star LMXB \source\ started a new, faint outburst in 2014 October. We obtained simultaneous \nustar\ and \swift\ observations during its initial hard X-ray spectral state. The continuum emission in the 3--79~keV \nustar\ band was dominated by non-thermal emission that can be described by a $\Gamma$$\simeq$2 power law, with a small ($\simeq$1--2 per cent) contribution from a soft emission component that can be modelled with an $\simeq$0.3--0.6~keV blackbody. The hard X-ray continuum and the excellent sensitivity of \nustar\ at energies $>$10~keV resulted in a high-quality disc reflection spectrum, showing both a broad Fe-K emission line peaking near 7~keV and a Compton hump around 20--30~keV. We fitted the spectral data with the versatile disc reflection model \relx\ to investigate the accretion geometry of \source, and to obtain constraints on the radius of the neutron star.

\subsection{Accretion geometry of \source}\label{subsec:incl}
It is generally thought that in soft X-ray spectral states, when the accretion rate is high ($\gtrsim$10 per cent of the Eddington limit), the accretion disc extends to/near the ISCO, whereas in quiescence the inner disc radius lies far from the compact object \citep[$\gtrsim$$100~\gmc$; e.g.,][]{esin1997}. Therefore it is expected that somewhere in between these extremes, in the hard X-ray spectral state, the inner edge of the disc should retreat. However, there is considerable debate whether the accretion disc is truncated throughout hard X-ray spectral states, or whether its inner radius only starts to recede below a certain accretion rate, because radii inferred from reflection and thermal emission components often yield opposing results \citep[e.g.,][]{rykoff2007,gierlinsky2008,tomsick2009,reis2010,done2010,kolehmainen2014,plant2014}. This discussion has mainly focused on black hole LMXBs. The accretion geometry could possibly be different in the hard X-ray spectral states of neutron star LMXBs, as the stellar surface and anchored magnetic field may come into play.

We estimate that \source\ accreted at $\simeq$1--2 per cent of the Eddington limit during our observations, which thus probe a relatively low accretion regime for disc reflection studies of neutron star LMXBs. Our spectral fits consistently point to an inner disc radius that lies close to the ISCO: $R_{\mathrm{in}}$$=$1.3--2.0$~\risco$. Given that $\risco$$\simeq$$5.05~GM/c^2$ for a neutron star spinning at $a$$=$$0.29$, this would correspond to $R_{\mathrm{in}}$$=$7--10$~\gmc$$=$15--21~km for a mass of $M$$=$$1.5~\Msun$ \citep[a reasonable choice based on the recent overview of][]{lattimer2014}. Our obtained inner disc radius for \source\ is within the range found for several other neutron star LMXBs \citep[$R_{\mathrm{in}}$$\simeq$5--20$~\gmc$; e.g.,][]{cackett2010_iron,egron2011}. Our study does not indicate significant truncation of the accretion disc, despite the low accretion rate \citep[see also][]{disalvo2015}. Due to uncertainties in our data at $\lesssim$3~keV, we cannot obtain reliable radius estimates from the thermal emission in \source. 

Our spectral analysis further points to a moderately low disc inclination angle of $i\simeq$30$^{\circ}$--40$^{\circ}$. This is consistent with the lack of dips/eclipses from \source\ and with the possible detection of a `superhump' \citep[][]{wachter2002}, which should be best observed at low inclination \citep[e.g.,][]{haswell2001}. Furthermore, we find that the illuminating hard X-ray source is likely located close to the neutron star, at a height of $h$$<$8.5~$\gmc$. This may be consistent with the growing consensus that the hard X-ray corona in similar accreting systems (black hole LMXBs and active galactic nuclei) is very compact \citep[e.g.,][for recent discussion]{reis2013_corona,fabian2014}. Alternatively, the very small height inferred from our reflection fits could point to the boundary layer between the accretion disc and the stellar surface as the primary source of the illuminating hard X-rays \citep[e.g.,][]{gierlinski2002}.

\subsection{Neutron star radius constraints}\label{subsec:inclination}
Since the disc must truncate at the surface of the neutron star if not at larger radii, reflection modelling can be used to place constraints on the neutron star radius \citep[e.g.,][]{cackett2010_iron,miller2013_serx1}. For a gravitational redshift of $1+z=1/\sqrt{1-2GM/Rc^2}$, the inner disc radius implied by our fits would constrain the neutron star radius to $R$$\lesssim $21~km, hence the gravitational redshift to $z$$\gtrsim$0.12 for an assumed mass of $M=1.5~\Msun$. These constraints from the disc reflection spectrum of \source\ are consistent with those obtained from its type-I X-ray bursts \citep{guver2010,poutanen2014_4u1608}.

\section*{Acknowledgements}
We thank the referee, Craig Heinke, for thoughtful comments. N.D. acknowledges support via an EU Marie Curie Intra-European fellowship under contract no. FP-PEOPLE-2013-IEF-627148. This work is based on data from the \nustar\ mission, a project led by California Institute of Technology, managed by the Jet Propulsion Laboratory, and funded by NASA. We thank Neil Gehrels and the Swift team for rapid scheduling of observations.

\footnotesize{
\bibliographystyle{mn2e}

}

\end{document}